# Multilevel recording in Bi-deficient Pt/BFO/SRO heterostructures based on ferroelectric resistive switching targeting high-density information storage in nonvolatile memories

**David Jiménez[a]), Enrique Miranda[a]), Atsushi Tsurumaki-Fukuchi,[b]) Hiroyuki Yamada[b),c]), Jordi Suñé[a]), and Akihito Sawa[b])**

[a])Departament d'Enginyeria Electrònica, Escola d'Enginyeria, Universitat Autònoma de Barcelona, 08193-Bellaterra, Barcelona, Spain

[b])National Institute of Advanced Industrial Science and Technology (AIST), Tsukuba, Ibaraki 305-8562 Japan

[c])JST, PRESTO, Kawaguchi, Saitama, 332-0012, Japan

**Abstract**

We demonstrate the feasibility of multilevel recording in $Pt/Bi_{1-\delta}FeO_3/SrRuO_3$ capacitors using the ferroelectric resistive switching phenomenon exhibited by the $Pt/Bi_{1-\delta}FeO_3$ interface. A tunable population of up and down ferroelectric domains able to modulate the Schottky barrier height at the $Pt/Bi_{1-\delta}FeO_3$ interface can be achieved by means of either a collection of SET/RESET voltages or current compliances. This programming scheme gives rise to well defined resistance states, which form the basis for a multilevel storage nonvolatile memory.

Ferroelectric materials with their electrically switchable spontaneous polarization have proven to be very effective in tuning the charge transport properties at a ferroelectric/metal interface[1-14]. A potential advantage over other mechanisms triggering the resistive switching (RS) is related to an intrinsic property of the material (ferroelectricity) rather than to extrinsic defects such as oxygen vacancies, whose concentration strongly depends on the particular features of the material. In addition, ferroelectric RS-based memories allow the possibility of non-destructive polarization state reading as compared with conventional ferroelectric RAMs (Fe-RAM) for which the read-out process deletes the stored logic state.

Multiferroic $BiFeO_3$ (BFO) displays interesting properties such as robust ferroelectricity[15] and small band gap[16] in comparison with other ferroelectric materials, ranging from 2.2 to 2.8 eV. The narrow band gap provides a relatively small Schottky barrier height at the metal/BFO contact that can be easily modulated by the large polarization charge of BFO (around 100 µC/cm^2 at room temperature along the [111] direction). Remarkably, this property has been used in Au/BFO/Pt capacitors, which show robust multilevel resistance states after irradiation of BFO with low-energy $Ar^+$ ions[17]. Recently, it has been reported that the performance of RS devices could be improved by controlling the amount of defect charges in BFO films. Specifically, the control of electronic transport by polarization reversal in a Bi-deficient Pt/BFO/$SrRuO_3$ (SRO) heterostructure epitaxially grown on a $SrTiO_3$ (STO) substrate was demonstrated[18,19], showing endurance of >$10^5$ cycles and data retention of >$10^5$ s at room temperature[18]. The Bi deficiency δ increases the valence of Fe ions and confers a p-type character to the $Bi_{1-\delta}FeO_3$ films. Our recent study on piezoresponse force microscope (PFM) in the Bi-deficient BFO devices demonstrated that the devices show the PFM phase hysteresis loops which well agree with the pulsed-voltage-induced RS hysteresis loops[20]. This result clearly indicates the ferroelectric origin of the RS, although the possibility of oxygen vacancies cannot be absolutely excluded.

In this paper, we further report on the tunability of the ferroelectric RS effect observed at a rectifying Pt/$Bi_{1-\delta}FeO_3$ interface. Here, the $Bi_{1-\delta}FeO_3$ film behaves as a p-type semiconductor acting as the switching element. Its conductivity, as well as the depletion layer width associated with the formation of a Schottky-like barrier, can be controlled by changing the Bi deficiency concentration δ. Remarkably, this structure does not require electroforming, which is advantageous respect to conventional RS in metal oxides.

The samples consisted in Au/Pt/BFO/SrRuO$_3$ (SRO) layered structures on SrTiO$_3$ (001) single-crystal substrates (Fig. 1a). A 50 nm-thick SRO bottom electrode was grown on the substrate prior to a 100 nm-thick BFO layer obtained by pulsed laser deposition (PLD). As revealed by X-ray diffraction measurements, both BFO and SRO layers were confirmed to be epitaxially grown on the SrTiO$_3$ substrates. An Au(100 nm)/Pt(10 nm) top electrode was deposited on the BFO layer through a shadow mask (pad size of 100 μm × 100 μm) by using electron-beam evaporation. In order to control the Bi content, the BFO films were deposited from source targets with controlled Bi/Fe ratio. In our experiments we focused on devices with the Bi/Fe ratio of 0.76 (± 0.05) estimated by inductively coupled plasma atomic emission spectroscopy. Details about the fabrication and characterization of such Bi-deficient BFO films, together with a confirmation of ferroelectricity of the conductive BFO layers by a piezoresponse force microscope (PFM) can be found in Ref. 18.

Current-voltage (*I-V*) characteristics were measured at room temperature using an Agilent 4155C semiconductor parameter analyzer. In all the measurements the top electrode was grounded and ramped voltages were applied to the bottom electrode (SRO). A typical *I-V* curve is shown in Fig. 1b (red line) upon application of a double-voltage sweep 5 V→ -5 V→ 5 V. Asymmetric bipolar-type switching with zero-crossing hysteretic characteristics are obtained. This behavior is ascribed to the resistive switching of a ferroelectric diode, where a Schottky-like potential barrier forms at the interface between the metal electrode and the conductive BFO. The potential profile of the barrier is reversibly modified by the polarization reversal[18]. Interestingly, the rectifying *I-V* characteristic inherently includes the functionality of the selector, i.e., a suitable rectifying element for avoiding crosstalk in crossbar patterned memory arrays[21,22]. *I-V* measurements are shown to be repeatable upon several cycles without significant modification (Fig. 1b, blue line). Measurements under illumination conditions were also performed. The increase of the measured current under light conditions respect to dark conditions, especially in the low-current regime at negative bias, corroborates the semiconductive nature of the BFO film (Fig. 2).

Next, we report on the RS tunability of the devices. First, we performed a series of double-voltage sweep ramps as -3 -> SET, where the SET voltage has been progressively increased from 2 to 3.5V. As seen in the left inset of Fig. 1b, the PFM phase hysteresis loop closes at a voltage between -3 V and -4 V in the negative voltage bias. The starting value of -3V, playing the role of the RESET voltage, thus provides a practically total domain reversal, but preventing a premature breakdown of the sample. In the experiment, the application of successive SET voltages provides an effective way to produce different amounts of domain switching. In fact, the gradual increase in the PFM phase from around 2.5 V in the positive bias (the left inset of Fig. 1b) supports a partial switching behavior of ferroelectric domains. The ferroelectric polarization modulates the Schottky barrier height at the $Pt/Bi_{1-\delta}FeO_3$ interface yielding multilevel resistance states. For instance, a nonvolatile memory of 7 states is demonstrated in Fig. 3a. The resistance states were read from the backward sweep at 1 V, spanning over two orders of magnitude (see inset). A schematic band diagram of the heterostructure is sketched in Fig. 3b, which graphically explains the barrier height modulation mechanism induced by the ferroelectric polarization as a function of the SET voltage. This band diagram arises from the modified Meyer's model to account for the Schottky barrier at the Pt/BFO interface[18,23]. Note that the band diagram includes interface layers (IL) formed between both contact electrodes and the BFO. The ferroelectric polarization (P) modulates the Schottky barrier height at the Pt/BFO interface, while the BFO/SRO interface behaves as an ohmic contact. Full downward and upward polarizations, shown as black and red arrows, respectively, produce either the low resistance or high resistance states, respectively. Multilevel resistance states can be prepared by inducing partial polarizations of the BFO shown as blue and green arrows. Because of the smaller Schottky barrier height at the valence band, estimated[18] to be less than 0.9 eV, as compared with the conduction band, the current is carried by holes, flowing from the SRO electrode to the Pt electrode.

Next, we performed a series of double-voltage sweeps as 5 -> RESET, where the RESET voltage was programmed to be 0,-1,-2,-3,-4, sequentially, while reading the resistance state along the forward sweep at 1 V. The results are shown in Fig. 4. A full RESET of the *I-V* characteristic is observed upon application of a bias down to -3V, corresponding to the HRS. A partial RESET is observed if the RESET voltage is greater than -3V. The strength of the RESET voltage can be used to tune the RS, so different memory states can be stored in a single device. Remarkably, it is feasible writing an arbitrary state without previously erasing the current state, which is certainly of utmost importance for fast memory operation. This can be seen by means of an experiment consisting of sequentially programming the states at 1V, 2V, 3V, 4V, 5V, and then reading at 1 V during the forward sweep, without resetting the memory between states. A similar experiment was done but programming non-ordered states at 2V, 5V, 3V, 4V, instead, and again reading at 1 V while performing the backward sweep. The resulting *I-V* characteristics are shown in Figs. 5a and 5b, respectively. The inset shows the resistance states at the reading voltage.

Tunability of resistance states is also possible by defining a set of compliance currents (Fig. 6), allowing to induce different amount of polarization in the BFO film. Here a double-voltage sweep as -4 -> 5V was applied to the sample under test combined with compliance currents of 10 nA, 100 nA, 1 μA, 10 μA, 0.1 A, respectively. Interestingly, we found a significant difference respect to the voltage controlled experiment shown in Fig. 3a. When the backward sweep is performed the current remains constant at the compliance level over an extra voltage beyond V* (see Fig. 6), producing a resistance state lower than it would correspond in a voltage controlled experiment with a SET voltage equal to V*. Here V* is defined as the crossing voltage between the forward and backward characteristics. Such $\Delta V$ increases with the strength of the compliance as shown in the inset. According to Lee et al.[24], the observation of multilevel RS could arise from the tuning of the speed of polarization switching by limiting the current flowing through the device. Setting an upper limit on the current, the speed of polarization switching can be exactly controlled, being the amount of the switched polarization proportional to the compliance current. Thus, by limiting the current, both control of the speed and amount of the polarization switching, would allow creating resistance states with any polarization value. However, this interpretation is limited to cases in which the displacement current dominates the whole conduction process. Additionally, this picture complicates even further because of the possible occurrence of the Maxwell-Wagner instability in dielectric stacks[25].

In summary, we have demonstrated multilevel tunability of resistance states of Bi deficient Pt/BFO/SRO heterostructures by using either the SET, RESET voltage, or the current compliance as well. We suggest that the mechanism behind is that of Schottky barrier height modulation provided by the fine level of control for the relative proportion of up and down domains, which is the basis for a multilevel storage ferroelectric nonvolatile memory.

**Acknowledgments**

This work was supported by the Ministerio of Economía y Competitividad of Spain under the project TEC2012-32305 (partially supported by the EU under the FEDER program). E.M. acknowledges the funding of the Agència de Gestió d'Ajuts Universitaris i de Recerca from Catalunya (CONES-2010). J.S. also thanks the funding support of the ICREA ACADEMIA award. and the DURSI of the Generalitat de Catalunya (2009SGR783).

**References**

[1]L. Esaki, R. B. Laibowitz, P. J. Stiles, *IBM Tech. Discl. Bull.* **13**, 2161 (1971).

P. W. M. Blom, R. M. Wolf, J. F. M. Cillessen, M. P. C. M. Krijn, *Phys. Rev. Lett* **73**, 2107 (1994)

[2]J. Rodriguez-Contreras, H. Kohlstedt, U. Poppe, R. Waser, C. Buchal, N. A. Pertsev, *Appl. Phys. Lett.* **83**, 4595–4597 (2003)

[3]V. Garcia, S. Fusil, K. Bouzehouane, S. Enouz-Vedrenne, N. D. Mathur, A. Barthélémy, M. Bibes, *Nature* **460**, 81–84 (2009).

[4]P. Maksymovych, S. Jesse, P. Yu, R. Ramesh, A. P. Baddorf, S. V. Kalinin, *Science* **324**, 1421–1425 (2009).

[5]A. Gruverman. D.Wu, H.Lu, Y.Wang, H.W.Jang, C.M.Folkman, M.Ye.Zhuravlev, D.Felker, M.Rzchowski, C.-B.Eom and E.Y.Tsymbal, *Nano Lett.* **9**, 3539–3543 (2009).

[6]A. Chanthbouala, A. Crassous, V. Garcia, K. Bouzehouane, S. Fusil, X. Moya, J. Allibe, B. Dlubak, J. Grollier, S. Xavier, C. Deranlot, A. Moshar, R. Proksch, N. D. Mathur, M. Bibes, A. Barthélémy, *Nature Nanotechnology* **7**, 101–104 (2012)

[7]A. Chanthbouala, V. Garcia, R. O. Cherifi, K. Bouzehouane, S. Fusil, X. Moya, S. Xavier, H. Yamada, C. Deranlot, N. D. Mathur, M. Bibes, A. Barthélémy, J. Grollier, *Nature Materials* **11**, 860-864 (2012).

[8]D. Lee, S. H. Baek, T. H. Kim, J.-G. Yoon, C. M. Folkman, C. B. Eom, T. W. Noh, *Phys. Rev. B* **84**, 125305 (2011)

[9]R.K. Pan, T.J. Zhang, J.Z. Wang, Z.J. Ma, J.Y. Wang, D.F. Wang, *J. Alloys and Compounds* **519**, 140– 143 (2012)

[10]L. Pintilie, I. Vrejoiu, D. Hesse, G. Lerhun, M. Alexe, *Phys. Rev. B* **75**, 104103 (2007)

[11]L. Pintilie, V. Stancu, L. Trupina, and I. Pintilie, *Phys. Rev. B* **82**, 085319 (2010)

[12]Z. Chen, L. He, F. Zhang, J. Jiang, J. Meng, B. Zhao, A. Jiang, *J. Appl. Phys.* **113**, 184106 (2013)

[13]D. J. Kim, H. Lu, S. Ryu, C. W. Bark, C. B. Eom, E. Y. Tsymbal, A. Gruverman, *Nano Lett.* **12**, 5697–5702 (2012)

[14]Z. Wen, C. Li, D. Wu, A. Li, N. Ming, *Nature Materials* **12**, 617-621 (2013)

[15]J. Wang, J. B. Neaton[2], H. Zheng, V. Nagarajan, S. B. Ogale, B. Liu, D. Viehland, V. Vaithyanathan, D. G. Schlom, U. V. Waghmare, N. A. Spaldin, K. M. Rabe, M. Wuttig, R. Ramesh, *Science* **299**, 1719 (2003)

[16]T. Choi, S. Lee, Y. J. Choi, V. Kiryukhin, S.-W. Cheong, *Science* **324**, 63 (2009)

[17]Y. Shuai, X. Ou, W. Luo, N. Du, C. Wu, W. Zhang, D. Bürger, C. Mayr, R. Schüffny, S. Zhou, M. Helm, H. Schmidt, *IEEE Electron Devices Lett.* **34**, 54-56 (2013)

[18]A. Tsurumaki, H. Yamada, A. Sawa, *Adv. Funct. Mater.* **22**, 1040-1047 (2012)

[19]T. H. Kim, B. C. Jeon, T. Min, S. M. Yang,. D. Lee, Y. S. Kim, S.-H. Baek, W. Saenrang, C.-B. Eom, T. K. Song, J.-G. Yoon, T. W. Noh, *Adv. Funct. Mater.* **22**, 4962–4968 (2012)

[20]A. Tsurumaki-Fukuchi, H Yamada, A. Sawa, Appl. Phys. Lett. ***103***, 152903 (2013)]

[21]E. Linn, R. Rosezin, C. Kügeler, R. Waser, *Nature Materials* **9**, 403-406 (2010)

[22]B. S. Kang, S. E. Ahn, M. J. Lee, G. Stefanovich, K. H. Kim, W. X. Xianyu, C. B. Lee, Y. S. Park, I. G. Beak, and B. H. Park, *Adv. Mater.* **20**, 3066-3069 (2008)

[23]R. Meyer, J. Rodriguez Contreras, A. Petraru, H. Kohlstedt, *Int. Ferroelectrics* **64**, 77-88 (2004) H. Kohlstedt, A. Petraru, K. Szot, A. Rüdiger, P. Meuffels, H. Haselier, R. Waser, V. Nagarajan, *Appl. Phys. Lett.* **92**, 062907 (2008)

[24]D. Lee, S. M. Yang, T. H. Kim, B. C. Jeon, Y. S. Kim, J.-G. Yoon, H. N. Lee, S. H. Baek, C. B. Eom, T. W. Noh, *Adv. Mater.* **24**, 402-406 (2011)

[25]J. R. Jameson, P. B. Griffin, J. D. Plummer, Y. Nishi, IEEE Trans. Electron Devices **53**, 1858-1867 (2006)

**Figure captions**

**Figure 1.** (a) Cross section of the Bi-deficient Pt/BFO/SRO heterostructure. (b) *I-V* characteristics upon application of seven consecutive double-sweep (DS) voltage ramp as -4->5 V (blue line) and a single double-sweep voltage ramp as -5->5 V (red line). Left inset shows out-of-plane PFM amplitude and phase hysteresis loops for the Bi-deficient Pt/BFO/SRO heterostructure. Right inset shows the measured ON/OFF current ratio. The dashed line indicates the optimum reading voltage.

**Figure 2.** *I-V* characteristics of Pt/BFO/SRO heterostructure under both light and dark conditions. The increase of the current under light illumination indicates that the BFO film behaves as a semiconductor.

**Figure 3.** Programming the resistance state with the SET voltage. (a) *I-V* characteristics demonstrating the feasibility of a 7-state nonvolatile memory. The inset shows the resistance state corresponding to a reading voltage of 1 V. (b) Schematic band diagram of Pt/BFO/SRO heterostructure at V>0.

**Figure 4.** Programming the resistance state with the RESET voltage: *I-V* characteristics corresponding to different memory states. The inset shows the resistance read at 1 V.

**Figure 5.** (a) Sequential writing and reading of the resistance states without erasing the current state. (b) Arbitrary writing and reading of the resistance states. Both insets of (a) and (b): resistance states read at 1 V.

**Figure 6.** Tunability of resistance states by using the compliance current (CC). Inset: extra voltage as a function of the CC.

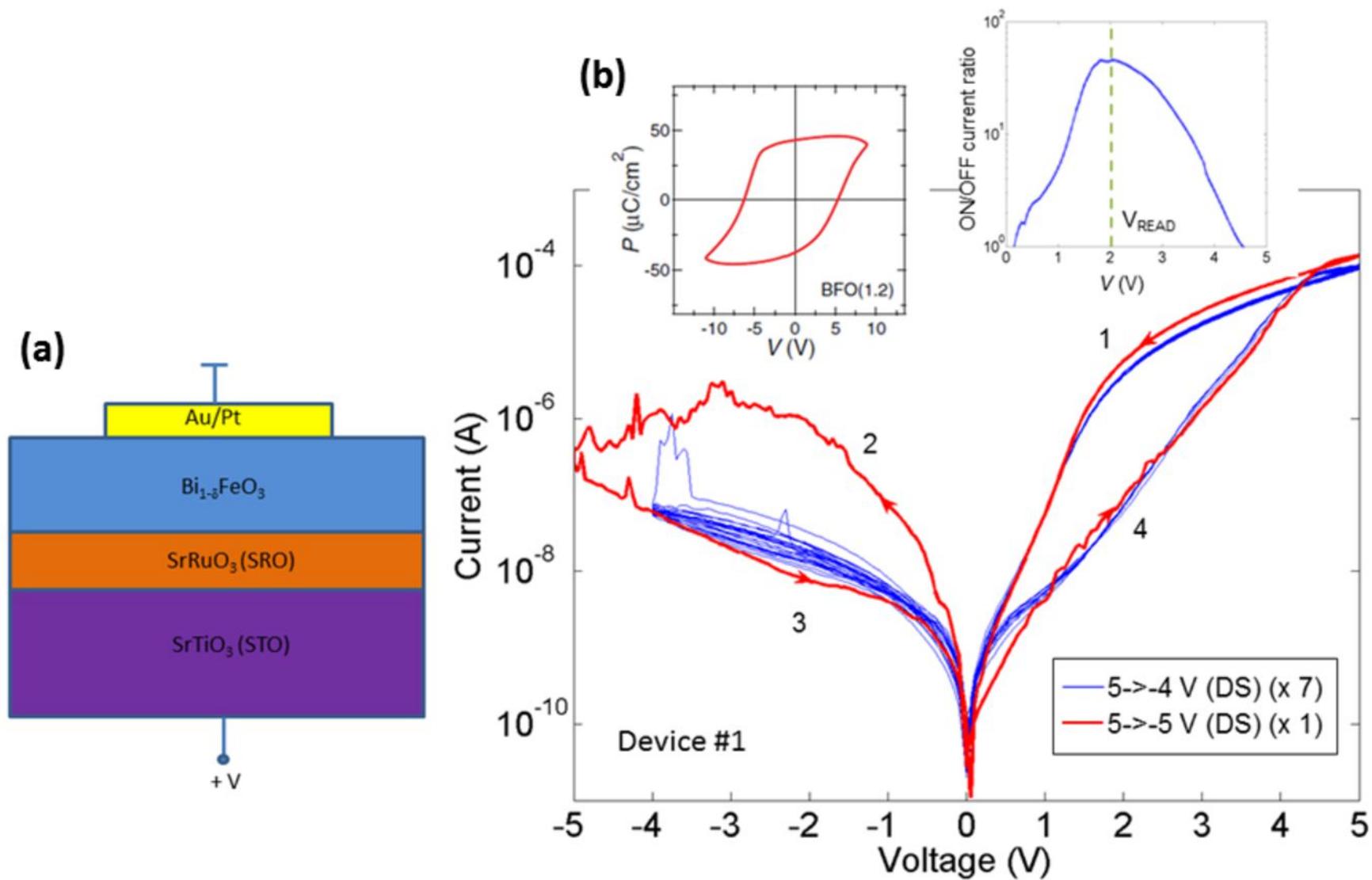

Figure 1

Figure 2

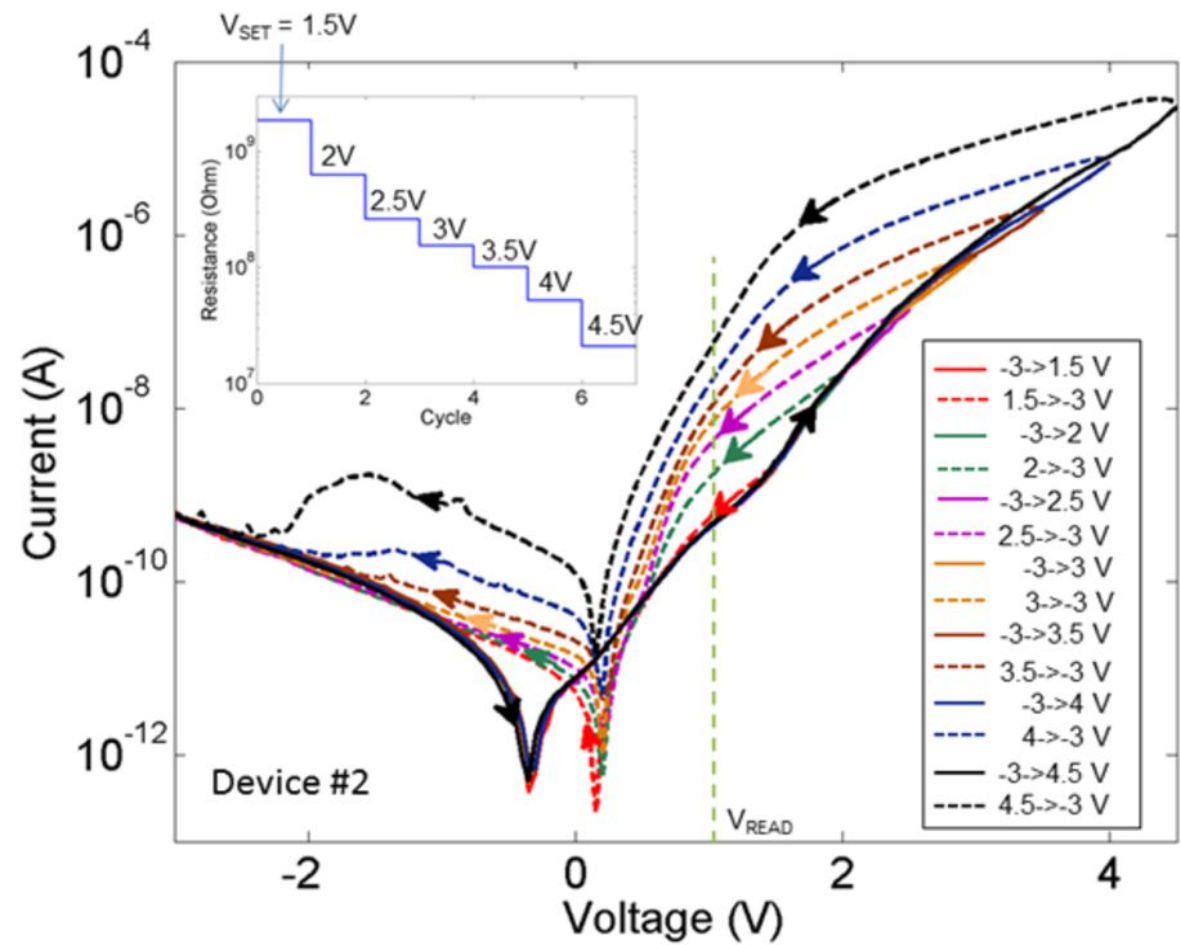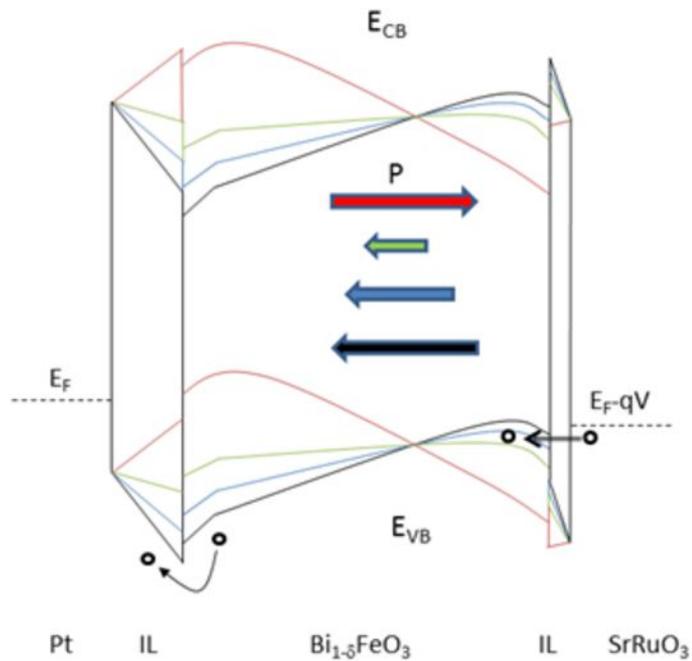

**Figure 3**

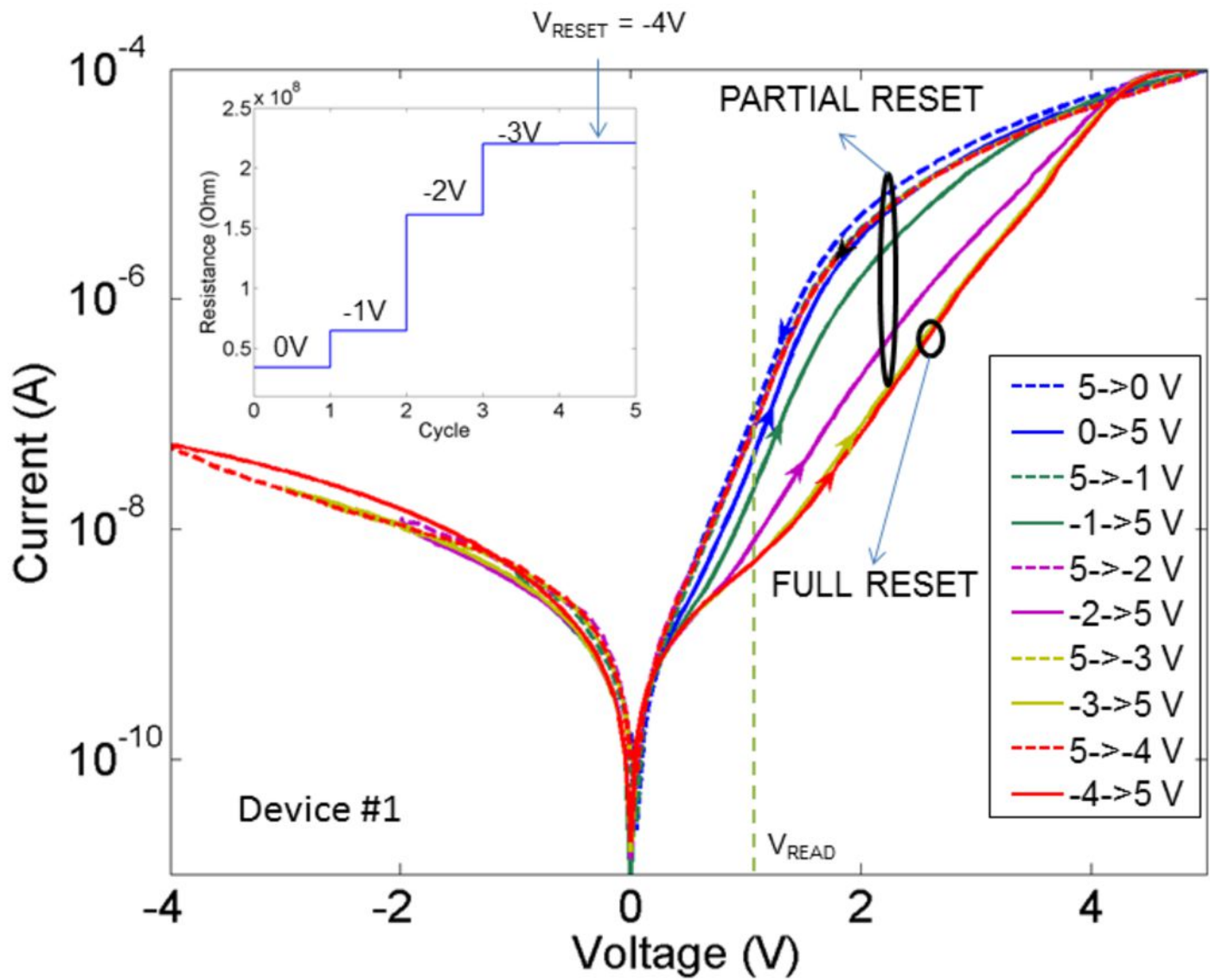

**Figure 4**

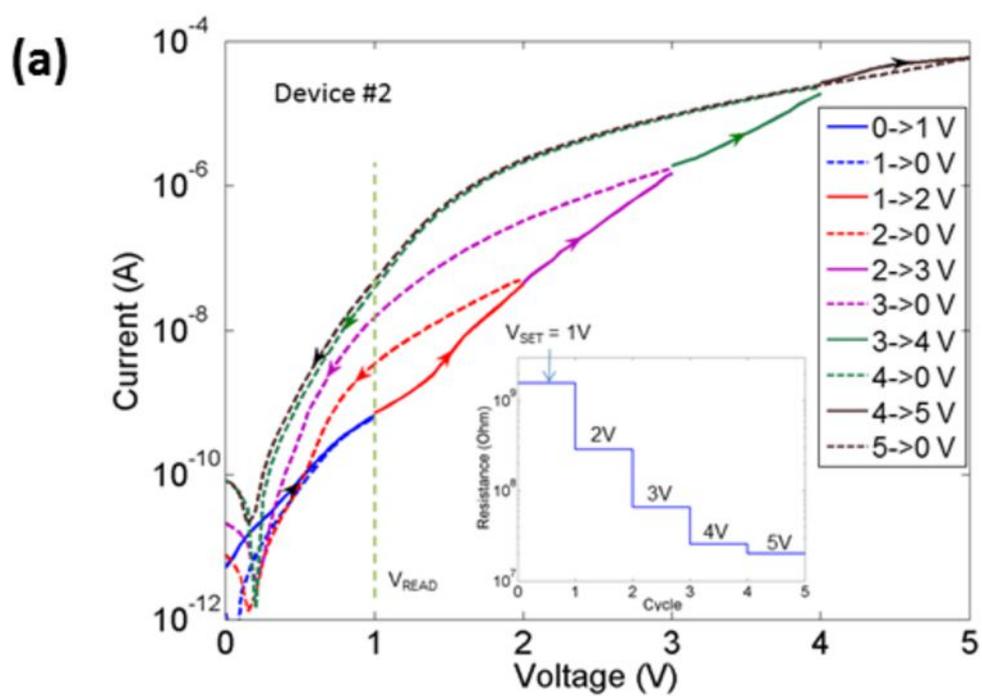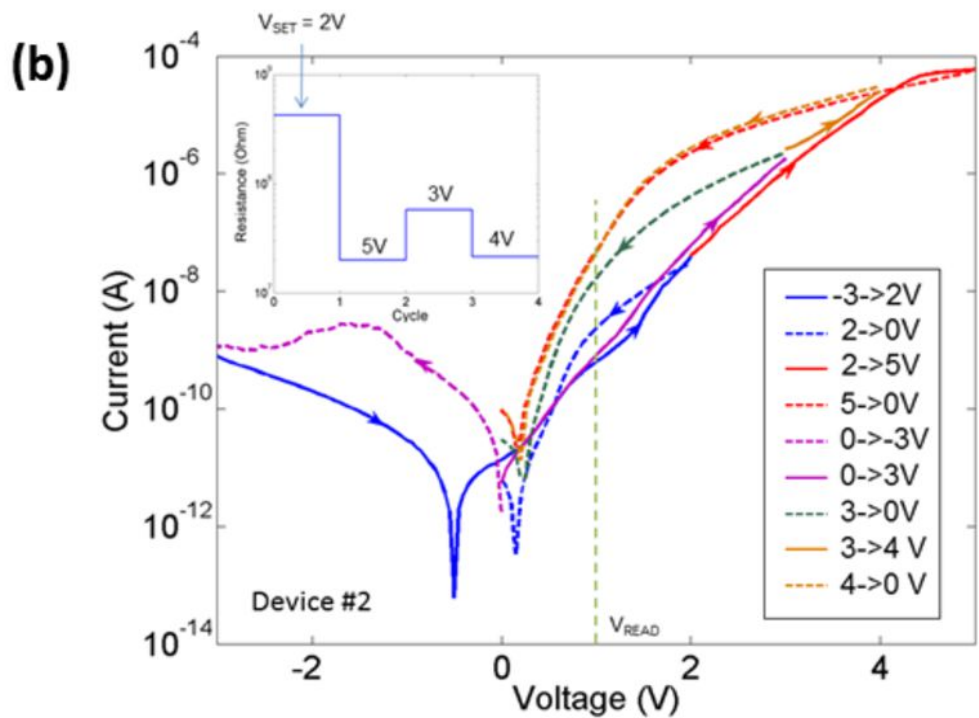

Figure 5

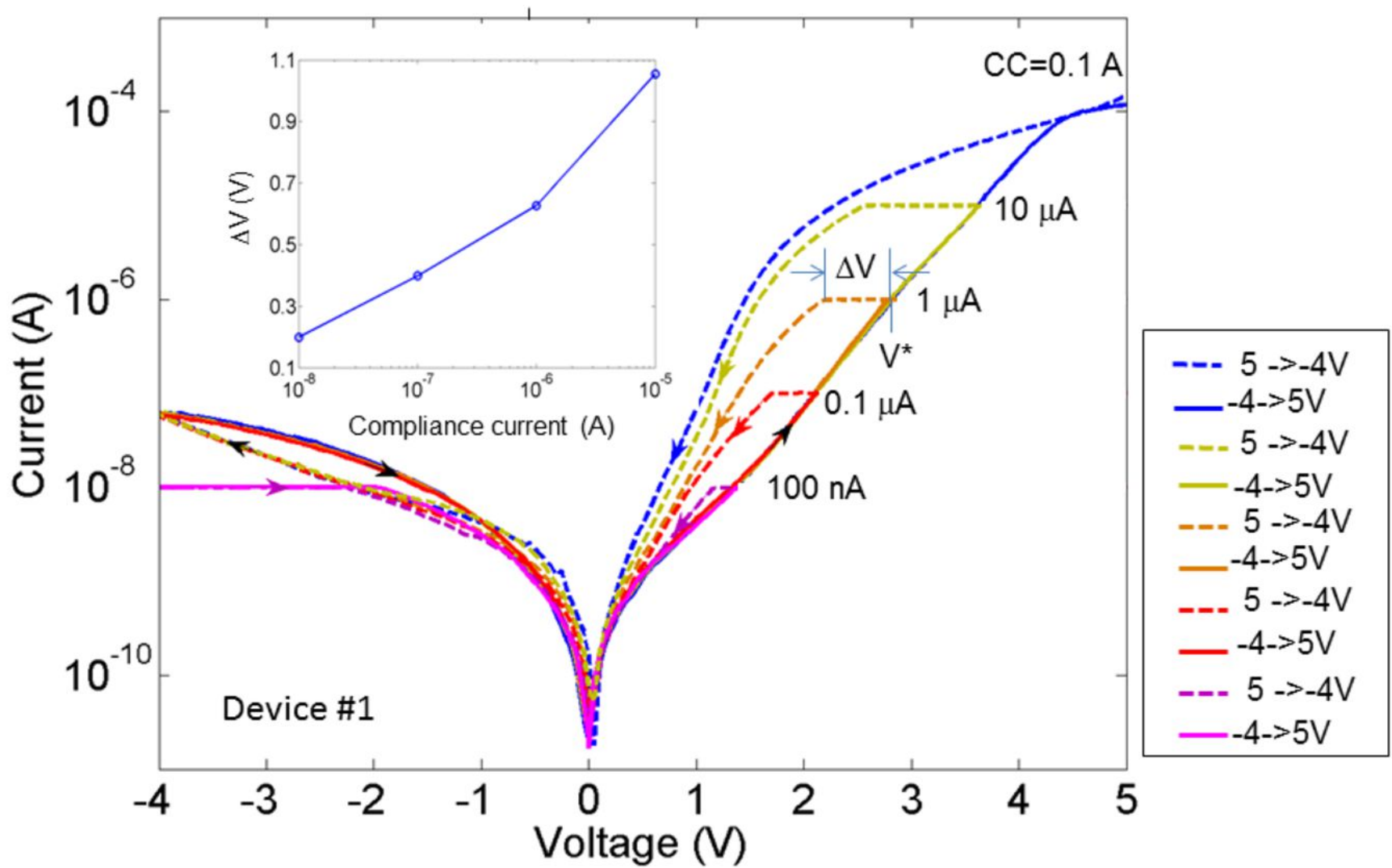

**Figure 6**